\documentstyle[epsf,prl,aps,twocolumn]{revtex}
\begin{document}
\draft
\title
{Binding energy of  static  perfect fluids.}
\author
{Janusz Karkowski   and Edward Malec}
\address{   Institute of Physics, Jagiellonian University,
30-059  Cracow, Reymonta 4, Poland.}
\maketitle
\begin{abstract}
We investigate the binding energy in two classes of polytropic perfect fluids.
A general-relativistic bound from below  is derived
in the case of a static  compact   body, having   the same form  as
in the newtonian limit.  It is shown  that   the  positivity of the
binding energy implies that the (properly defined) average speed of
sound is smaller than the escape velocity. A necessary condition
for the negative binding energy, stating that the maximal speed of sound
is close to the escape velocity, is found in a class of fluids.

\end{abstract}

\section{Introduction}

There are two popular classes of equations of states of perfect fluids.
One, that is commonly used by astrophysicists,  takes a simple form when
formulated   in terms of the so-called baryonic
number density. The other class, preferred by general relativists, uses
the mass density.  The two classes coincide in the newtonian
limit but they can lead to markedly different predictions in the strong
field regime (see, for instance, a comparative discussion of stationary accretion in \cite{bogusz}).
It is intention of this paper to rederive some known results
and obtain new ones  on the binding energy of perfect fluids
satisfying polytropic equations of state of the two types. A   possible new feature
of the derivation is that we do not recourse to the laws of
relativistic thermodynamics. We restrict our attention to static fluid bodies,
but since we operate in those  classes of equations  of state for which
staticity implies spherical symmetry  \cite{Beig},  we end
in investigating  spherically symmetric   objects.

The order of this paper is as follows.   Section 2 presents equations of selfgravitating fluids.
In the third section we recall definitions of the two classes of relativistic
equations of state and show that in the (properly defined) Newtonian limit they coincide.
 Section 4  defines the gravitational binding energy and in Secs 5    its value
 is estimated in the polytropic subclass  of the equations of state.
In the newtonian limit the binding energy agrees with the standard newtonian expression.
Section 6 is dedicated to the derivation of certain inequalities onto the
gravitational potential energy. It is remarkable that these  inequalities  are
exact, and one of them has the same form both in the general-relativistic
and in the newtonian cases. Section 7 presents the main results.
It is shown that the
binding energy of a compact spherical body is bounded from below by the
standard newtonian  formula $-P/(\Gamma -1) + 3m^2_{TB}/(5R)$. The kinetic energy of
the gas is represented by the total pressure $P$ and the sufficient condition for having
a positive value of the binding energy is that $P$ is smaller than the
gravitational potential type term $3m^2_{TB}/(5R)$.  That can be translated
into a bound from above onto a quantity that represents
a kind of an average speed of sound -- that should not exceed the escape velocity
(or even the first cosmic velocity). A necessary condition (a lower bound
on the maximal speed of sound) for the negative binding
energy is formulated in the so-called $\tilde p - \rho $ class of fluids.

\section{Equations for selfgravitating fluids.}

Throughout this paper the Newtonian constant $G$ and
the velocity of light $c$ are put equal to 1. Let the selfgravitating system be
spherically symmetric. Then the whole dynamics is carried by a material field
characterized by an energy-momentum tensor $T_{\mu \nu}$. In the case of the perfect
fluid $T_{\mu \nu }=(\tilde p+\rho )U_{\mu }U_{\nu }+\tilde pg_{\mu \nu }$, where the
coordinate velocity is timelike, $U_{\mu }U^{\mu }=-1$.

Assuming   spherical symmetry, one can deal with the metric
\begin{equation}
ds^2=-N^2dt^2+\alpha dr^2+R^2d\Omega^2,
\label{0}
\end{equation}
where $N$, $R$ and $\alpha $ are   unknown metric functions. In sections
2, 3 and in the first part of Sec. 4 we work in comoving coordinates. Our
aim there is to justify a particular expression of the binding energy, and
that requires  some information about the evolution that led to the
formation of a bound system. Comoving coordinates are convenient in describing
evolving  fluids. Starting from section 4 we use standard (polar gauge) coordinates.

The needed material quantities
are the energy density  $\rho =-T_0^0$ and the isotropic pressure,
$\tilde p=T_r^r=T_{\theta }^{\theta }=T_{\phi }^{\phi }$.
Define $pR\equiv 2\partial_rR/\sqrt{\alpha }$ (this is so-called
mean curvature of a centered sphere \cite{mom}) and $U\equiv \partial_tR/N$.
One can show, using the constraint equations \cite{malec99} and the properties of
comoving coordinates, that
\begin{equation}
pR=2\sqrt{1-2{m(R)\over R}+U^2},
\label{1.2}
\end{equation}
where $m(R)$ is the quasilocal mass specified below.
The trace of the extrinsic curvature
\begin{equation}
trK= K^i_i\equiv
{1\over N}\partial_0 \ln \sqrt{\det (g^3_{ij})}
\label{1.2a}
\end{equation}
can be found from the momentum constraint
 \cite{malec99},

\begin{equation}
trK=2{\partial_r(UR^2)\over \sqrt{a}R^3p}.
\label{1.3}
\end{equation}
The continuity equations $\nabla_iT^i_j=0$ reduce to
two equations
\begin{equation}
  N \partial_r \tilde p
 +  \partial_rN (\tilde p+\rho)=0,
\label{1.4}
\end{equation}
\begin{equation}
\partial_0  \rho=
 -N tr K (\tilde p +\rho ).
\label{1.5}
\end{equation}
The Einstein evolution equations reduce to the single equation
\begin{equation}
\partial_0U= -N{m\over R^2} +{\partial_rR\partial_rN\over \alpha }
-4\pi NR\tilde p.
\label{1.6}
\end{equation}
 The quasilocal mass ("mass function")
\begin{equation}
m(r)= \int_{V(r)} dV{Rp \over 2}\rho
\label{1}
\end{equation}
contained within a coordinate sphere $r'\le r$
satisfies  the evolution law
\begin{equation}
\partial_0 m( r ) =
- 4\pi  N R^2U \tilde p.
\label{1.1}
\end{equation}

The mass $m(r)$ (which happens to be equal to the volume
representation of the Hawking mass \cite{mom}) can be expressed also as
\begin{equation}
m(R(r)) = 4\pi  \int_0^{R(r)}dr r^2\rho ;
\label{1.7}
\end{equation}
the integrand is now written in terms of the areal radius. The total
mass is conserved; the asymptotic mass  $m(\infty )$ is constant
on the whole foliation that develops from a given initial hypersurface.
There exists another mass measure, known as the rest
mass \cite{Wald},
\begin{equation}
M(R)= \int_{V(R)}dV\rho = 8\pi \int_0^Rdr {r \over p}\rho .
\label{1.8}
\end{equation}
The rest mass $M(\infty )$ is not conserved -- it might change from a slice
to a slice  (that is with time). In what follows we will
 write $M$ and $m$ instead of $M(\infty )$ and $m(\infty )$

The baryonic mass (the mass density $n$ is defined in the
next section)  contained within a coordinate sphere $r$,
\begin{equation}
m_{bar}(r)= \int_{V(r)}dVn = 8\pi \int_0^{R(r)}dr {r \over p}n ,
\label{1.9}
\end{equation}
is conserved: $\partial_0 m_{bar}(r)=0$.    That follows easily from
equations (\ref{0d}), (\ref{1.5}) and  (\ref{1.2a}).   Yet another   mass
characteristic  of a static  body comprised within the coordinate sphere $r$  is
its Trautman-Bondi mass $m_{TB}$, equal to    the quasilocal mass function
$m(r)$. This mass is equal to the asymptotic mass minus that fraction of
the fluid that has been dispersed away to infinity. Therefore $m_{TB}\le m$.

\section{Equations of state}

The ($\tilde p -n $ henceforth) equation of state used by astrophysicists is
\begin{equation}
\tilde p=\tilde p(n)
\label{0a}
\end{equation}
while the community of general relativists prefers
($\tilde p -\rho $ thereafter)
\begin{equation}
\tilde p=\tilde p(\rho ) .
\label{0b}
\end{equation}
Here $n$ is the baryonic mass density that can be defined
(up to a normalization factor)
as an integrability factor that ensures the conservation of
the baryonic current
\begin{equation}
j^{\mu }=nU^{\mu } .
\label{0c}
\end{equation}
One can easily show \cite{malec99} that the conservation laws
$\nabla_{\mu }T^{\mu }_{\nu }=0$
imply, under conditions assumed above, the relation
\begin{equation}
n=n_0\exp \Biggl( \int_{\rho_0}^{\rho }ds{1\over s+\tilde p(s)}\Biggr) .
\label{0d}
\end{equation}
Here $n_0$ and $\rho_0$ are constants that can be used in order to suitably normalize $n$.
Thus, given $\rho $ and an equation of state $\tilde p =\tilde p(\rho )$   satisfying
$\nabla_{\mu }T^{\mu }_{\nu }=0$, one  finds $n$ automatically satisfying the
continuity equation $\nabla_{\mu }(nU^{\mu })=0$.

And conversely, given conserved baryonic density and an equation of state
 $\tilde p = \tilde p(n)$  one can find $\rho $,
\begin{equation}
\rho =n   \int_{ 0}^{n }ds{\tilde p(s) \over  s^2} +n,
\label{0e}
\end{equation}
such that $\nabla_{\mu }T^{\mu }_{\nu }=0$.
The formulae are particularly simple in the case of polytropic equations of state. Assuming
\begin{equation}
 \tilde p = K\rho^{\Gamma }
\label{0f}
\end{equation}
with $\Gamma >1$ one has
\begin{equation}
n=n_0{\rho \over (1+{a^2\over \Gamma })^{1\over \Gamma -1}}  
\label{0g}
\end{equation}
($a$ is the speed of sound).
If one assumes $n_0=1$ then, defining the  newtonian limit to be  the case $a^2\ll 1$,
it follows that in the newtonian limit the baryonic density and the mass density are very close,
$n\approx \rho $.

Assuming the alternative state equation,
\begin{equation}
\tilde p=Kn^{\Gamma },
\label{0h}
\end{equation}
 one obtains from Eq. (\ref{0e})
\begin{equation}
\rho = n +{\tilde p\over \Gamma -1};
\label{0i}
\end{equation}
if the newtonian limit is understood as the  stronger than before  condition  $\tilde p/(\Gamma -1)\ll 1 $, then
$\partial_{\rho }n\approx 1$ and again $n\approx \rho $. Assuming in Eq. (\ref{0g}) that $n_0\rightarrow 0$ and
simultaneously keeping the pressure $\tilde p$ fixed (that is $Kn_0^{\Gamma }=const$), one obtains
the Harrison et al. equation of state \cite{harrison}
\begin{equation}
\rho =  {\tilde p\over \Gamma -1}.
\label{0j}
\end{equation}

In the following part of this work we shall deal with equations of state given by
(\ref{0f})   and   (\ref{0h}).  The adiabatic index $\Gamma $ is assumed to be strictly bigger
than 1. While the above relations   are  well known, it is interesting that
the above derivation can be done solely within classical general relativity,
without appealing to thermodynamics.

\section{Binding energy}

The binding energy $E$ of a static system  $S$ can be defined as the difference
between the initial mass that is needeed in order to create $S$ and the final
Trautman-Bondi mass. The most natural definition  would
be to put $E=m-m_{TB}$; $E$ is the difference between  the asymptotic (ADM)
mass $m$ and the mass $m_{TB}$ of the  final product. Traditionally one defines
the binding energy somewhat differently, using the concept of the baryonic
mass.  We will adhere to the standard approach, defining another
measure $E_{bind}$ of the binding energy; the reader may check later
that $E\ge E_{bind}$.

In the case of initial mass distributions that are at rest and  so
dispersed that the curvature of the Cauchy hypersurface is {\it practically} negligible,
the  two masses $m$ and  $M$  can be {\it practically} equal. If we assume that
matter consists exclusively of baryons, then (again, under the   condition that
the fluid is so diluted    that its pressure is negligible)  the baryonic mass $m_{bar}(r)$
and the mass function   $m(r)$ can be regarded as being equal initially.
Since  usually only a fraction of the gas (say, comprised within the coordinate sphere $r$)
becomes solidified in a form of the static object $S$, and it is the baryonic mass
$m_{bar}(r)$ that remains constant during a collapse, it is justifiable  to
define the binding energy $E$ as \cite{tooper}
\begin{equation}
E_{bind}=m_{bar}(r)-m_{TB}.
\label{2.1}
\end{equation}
Let us point out that the baryonic mass can be expressed -- using formulae of Sec. 3 -- in terms
of quantities describing the static system $S$. From now on we will employ standard (Landau-Lifschitz
\cite{Landau} ) coordinates with the radial coordinate being always the areal radius.
This implies $\alpha =1/ (1- {2m(R)\over R})$, where the mass function is given by
Eq. (\ref{1.7}).  All volume integrals will be performed over the volume $V$ of $S$.
In order to simplify notation, the total rest mass $M(r)$  of $S$ will  be denoted by
$M_S$ and the binding baryonic mass $m_{bar}(r)$ will be written simply as $m_{bar}$.
Notice that $m_{TB}$ is this quantity that enters the newtonian formula
describing the late-time
interaction of $S$ with a distant (but  located at a  finite distance) test body.

Let us define total pressure in $S$ by
\begin{equation}
P=\int_{V }dV\tilde p = 8\pi \int_0^Rdr {r \over p}\tilde p.
\label{2.2}
\end{equation}
It is convenient to rewrite formula (\ref{2.1}) in two different forms,
\begin{equation}
E_{bind}=M_S-{P\over \Gamma -1}-m_{TB}.
\label{2.3}
\end{equation}
in the case of the $ \tilde p -  n$ equation of state and
\begin{equation}
E_{bind}=\int_{V }dV {\rho \over  (1+{a^2\over \Gamma })^{1\over \Gamma -1}} -m_{TB}.
\label{2.4}
\end{equation}
$ \tilde p - \rho  $ equation of state. The equivalence  of (\ref{2.4}) and  (\ref{2.3})
with (\ref{2.1}) follows from (\ref{0i}) and (\ref{0g}), respectively.

In the newtonian limit  both expressions yield the same quantity
\begin{equation}
E_{bind}=\int_{V }dV \Bigl( {\rho m(r) \over  r} -{\tilde p\over  \Gamma -1}\Bigr) .
\label{2.5}
\end{equation}

\section{Estimates of the binding energy  }

One can invoke to  a workable version of the virial theorem, that exists in
static spherically symmetric systems, in order to eliminate $P$ from
(\ref{2.3}) (\cite{Simon} - \cite{georgiou}). This is obtained from
the Oppenheimer-Volkov equation
\begin{equation}
\Bigl( \tilde p+\rho  \Bigr) \Bigl( {m(R)\over R^2}+4\pi \tilde p R\Bigr) =
-\partial_R\tilde p \Bigl( 1- {2m(R)\over R}\Bigr) ;
\label{3.1}
\end{equation}
Multiplying (\ref{3.1}) by $4\pi R^3/\sqrt{ 1- {2m(r)\over r}}$ and integrating with
respect $R$, one gets (integrating by
parts, in order to eliminate the grad-term $\partial_R\tilde p$)
\begin{eqnarray}
&&3\tilde P\equiv 3\int_VdV \tilde p \left( 1- {2m(R)\over R}\right) = \nonumber \\
&& -U_G+4\pi \int_VdV\tilde p  r^2 \left( \tilde p +2\rho \right) .
\label{3.2}
\end{eqnarray}
Here $U_G$ is the gravitational self-interaction energy,
\begin{equation}
U_G\equiv - \int_VdV     {\rho  m(r)\over r }.
\label{3.2a}
\end{equation}
The total pressure $P=\int_{V}dV \tilde p$ is not smaller than $\tilde P$, as one can see
rewriting the first line of (\ref{3.2})
\begin{equation}
P=\tilde P +\int_VdV {2m(r)\tilde p\over r}.
\label{3.2b}
\end{equation}
The mass difference  $M_S-m_{TB}$   reads (this follows directly
from definitions of the two  masses)
\begin{equation}
M_S-m_{TB}=-\tilde U_G=\int_VdV      {2 \rho m(r)\over r\left( 1 +\sqrt{1- {2m(r)\over r}}
\right) }.
 \label{3.2c}
 \end{equation}
Define
\begin{equation}
-\hat U_G=-\tilde U_G +{1\over 3\Gamma -4}
\int_VdV {2 \rho m(r)^2\over r^2\left( 1 +\sqrt{1- {2m(r)\over r} } \right)^2}.
 \label{3.2d}
 \end{equation}
Employing the above information,  after some calculations one arrives    at
\begin{eqnarray}
 &&M_S-{P\over \Gamma -1}-m_{TB} =
\nonumber \\
&&M_S-{\tilde P\over \Gamma -1}-m_{TB}
-{1\over  \Gamma -1 }\int_VdV {2m(r)\tilde p\over r}=
\nonumber \\
&&- {3\Gamma -4\over 3\left( \Gamma -1\right) }\hat  U_G
 -{1\over \left( \Gamma -1\right) }\int_VdV {2m(r)\tilde p\over r} -
\nonumber \\
&&{4\pi \over 3\left( \Gamma -1\right) }\int_VdV      \tilde p r^2
\left( \tilde p +2\rho \right)    .
\label{3.3}
\end{eqnarray}
  Thence,
\begin{eqnarray}
 M_S-{P\over \Gamma -1}-m_{TB}\le  - { 3\Gamma - 4\over 3(\Gamma -1)}\hat U_G.
\label{3.3a}
\end{eqnarray}
Notice, that $E_{bind}=   M_S-{P\over \Gamma -1}-m_{TB}$
for the $\tilde p-n$ equation of state. Therefore  we infer that
the binding energy may become negative for values of the polytropic index
$  \Gamma \le 4/3$. In the newtonian limit one obtains
\begin{eqnarray}
E_{bind}=-{ 3\Gamma - 4\over 3(\Gamma -1)}U_G.
\label{3.4}
\end{eqnarray}
In the newtonian limit the binding energy becomes strictly negative only when
$\Gamma <4/3$, while in the regime of strong gravitational fields this may happen
even for $\Gamma =\epsilon +4/3$, for small but positive $\epsilon $.
The general relativistic effects may decrease the binding energy (in agreement with
\cite{tooper}).
Let us remark here, that the result in \cite{tooper} corresponding to Eq. (\ref{3.4})
has a different coefficient (for reasons not quite clear to us):
 $(3\Gamma -4)/(5\Gamma -6)$ instead of $(3\Gamma - 4)/(3(\Gamma -1))$.

Notice that  always  $E_{bind}<M_S -m_{TB}$.  In the nonrelativistic case and putting
$ \Gamma = 5/3$ one
obtains  a stronger result, $E_{bind}=- U_G/2\approx (M-m_{TB})/2$; the quantity $M_S-m_B$, that
can be regarded as the gravitational potential energy in the case of static systems
(or perhaps even momentarily static systems), bounds the binding energy  and it is
of the order of the binding energy. The second part of the last  statement  is not
valid for  strongly curved geometries, even in the case $\Gamma =5/3$.

Assuming the polytropic $\tilde p -\rho $ equation of state,  we can express the binding energy
by formula (\ref{2.4}). It is easy to show
that  $\left( {1\over 1+ {a^2\over \Gamma }}\right) ^{1\over \Gamma -1} \ge
 1 - {a^2\over \Gamma  (\Gamma -1)}$.
Therefore one arrives at
\begin{eqnarray}
E_{bind}\ge  M_S - {P \over \Gamma -1}-m_{TB}.
\label{4.2}
\end{eqnarray}
It is clear that now the use of the   virial theorem  (which  bounds from
above  $M_S - {P \over \Gamma -1}-m_{TB}$) cannot give an estimate from
above on the binding energy.
There exists only a trivial bound $E_{bind}\le -\tilde U_G$ (this is because
$M_S\ge m_{bar}$) -- the binding energy is bounded by the absolute value of
the gravitational potential energy.
One can see,   comparing formulae (\ref{3.3}) and (\ref{4.2}), that the $\tilde p -\rho $
equation of state  is more favourable for having nonnegative binding energy that the $\tilde p - n$ case.
Nevertheless, in the newtonian limit one again
recovers estimate (\ref{3.4}). We will show in the last section that by exploiting
the quantity $M-m_{TB}$ one can find another  useful bound from below.

\section{Bounding potential energy}

In this section one might partly relax  former conditions, and
to deal  with a compact   (not necessarily fluid) body on the momentarily
stationary spatial 3-dimensional geometry.
The    assumptions i)  of staticity, and  ii) that  matter is a  barotropic fluid,
$\tilde p = K \rho^{\Gamma (\rho )}$,
become relevant in a refined version of the forthcoming result.
To be specific, the second and third  terms on the right hand side of Eq. (\ref{5.2a}) would be absent
without the two last conditions. Let us point that the compactness of the fluid body
imposes a bound from below on the barotropic index (\cite{walter} and
\cite{Heinzle}).

It is easy to show the following result, that estimates the difference
$M_S -m_{TB} $.

{\bf Theorem.} Assume that  matter
has compact support enclosed within a sphere $R$, that no minimal surface does exist
on the Cauchy slice and  $\rho \ge 0$.   Then
\begin{eqnarray}
-\tilde U_G & =& M_S-m_{TB}\ge {m^2_{TB}\over R}{2\over \Bigl( 1+\sqrt{1-{2m_{TB}\over R}}\Bigr)^2}
 \ge  \nonumber \\
&& {m^2_{TB}\over 2R}.
\label{5.2}
\end{eqnarray}
If in addition  the configuration is static, then
\begin{eqnarray}
-\tilde U_G & = &M_S-m_{TB}\ge
{m^2_{TB}\over R}{2\over \Bigl( 1+\sqrt{1-{2m_{TB}\over R}}\Bigr)^2}-
\nonumber\\
&&
{m_{TB} \over 2\left( 1+\sqrt{1-{2m_{TB}\over R}}\right) }+ \nonumber\\
&&{3R\over 4}
\left( -1+\sqrt{R\over 2m_{TB}}\arcsin (\sqrt{2m_{TB}\over R})\right) \ge
\nonumber\\
&& {3m^2_{TB}\over 5R}.
\label{5.2a}
\end{eqnarray}
{\bf Proof of the theorem.} Straightforward calculation allows one to check that
\begin{eqnarray}
&&
M_S = {1\over 2} \int_0^Rdr \Biggl( -2\sqrt{r}{d\over dr}\sqrt{r-2m(r)} +{ 1\over
\sqrt{1-{2m(r)\over r}}}
\Biggr) =\nonumber
\\
&&-R\sqrt{1-{2m_{TB}\over R}} +{1\over 2}\int_0^Rdr \Bigl( {1\over \sqrt{1-{2m\over r}}} +\sqrt{1-
{2m(r)\over r}} \Bigr) .
\label{5.3}
\end{eqnarray}
The integrand of the last integral is bounded from below by 2.
Therefore $M_S\ge R(1-\sqrt{1-{2m_{TB}\over R}})$; thus
\begin{eqnarray}
&&
M_S-m_{TB}\ge R(1-\sqrt{1-{2m_{TB}\over R}})-m_{TB}
=
\nonumber \\
&&{m^2_{TB}\over R}{2\over \Bigl( 1+\sqrt{1-{2m_{TB}\over R}}\Bigr)^2} \ge
\nonumber\\
&&{m^2_{TB}\over 2R}.
\label{5.4}
\end{eqnarray}
We can get only thus far in the   case of momentarily stationary data.

If our configuration is static, then the Oppenheimer-Volkov equation implies
$\partial_R\rho =a^{-2}\partial_R\tilde p\le 0 $. This in turn yields $\partial_r(m(r)/r^3)\le 0$
(that result was originally derived by Buchdahl \cite{Buchdahl}).

Returning to Eq. (\ref{5.3}), notice that
\begin{eqnarray}
 &&{1\over 2}\int_0^Rdr
\Bigl( {1\over \sqrt{1-{2m\over r}}} +\sqrt{1-{2m(r)\over r}} \Bigr)  =
\nonumber \\
&&
R+ \int_0^R{dr\over r^2} {2m^2\over  \sqrt{1-{2m\over r}}\left( 1+\sqrt{1-{2m\over r}}\right)^2}.
\label{5.4a}
\end{eqnarray}
Since $m(r)/r^3$ is decreasing, we have $m(r)/r^3 \ge m_{TB}/R^3$.
 Therefore
\begin{eqnarray}
&&2\int_0^R{dr\over r^2} {m^2\over  \sqrt{1-{2m\over r}}\left( 1+\sqrt{1-{2m\over r}}\right)^2}\ge
\nonumber\\
&&{2m^2_{TB}\over R^6}\int_0^Rdr { r^4  \over  \sqrt{1-{2m_{TB}r^2\over R^3}}
\left( 1+\sqrt{1-{2m_{TB}r^2\over R^3}}\right)^2 } .
\label{5.5}
\end{eqnarray}
The second integral term in Eq. (\ref{5.5}) can be explicitly evaluated, with the result
\begin{eqnarray}
F&\equiv &{-m_{TB} \over 2\left( 1+\sqrt{1-{2m_{TB}\over R}}\right) }+
\nonumber\\
&& {3R\over 4}
\left( -1+\sqrt{R\over 2m_{TB}}\arcsin \sqrt{2m_{TB}\over R}\right) .
\label{5.5a}
\end{eqnarray}
Combining  (\ref{5.3}), (\ref{5.5}) and (\ref{5.5a}), one gets the
first of the sought inequalities of  (\ref{5.2a}).  The second inequality there follows
from the trivial estimate $4\int_0^Rdr { r^4  \over  \sqrt{1-{2m_{TB}r^2\over R^3}}
\left( 1+\sqrt{1-{2m_{TB}r^2\over R^3}}\right)^2 }\ge R^5/5$.
That accomplishes the proof of the theorem.

Let us again stress  here that while in   (\ref{5.2}) one  needs only that initial
data are momentarily static,  the result with the coefficient $3/5$ in
 (\ref{5.2a}) requires staticity
(without that we would have only the coefficient $1/2$).  The two results  are exact.
The first  inequality is saturated by the spherical shell of matter
(see an example studied in \cite{bmom}) while the  inequalities of
(\ref{5.2a}) are saturated both in the general-relativistic and the
newtonian cases, by a constant density star.     In the case of fluid bodies
satisfying the Buchdahl limit $R=9m_{TB}/4$ \cite{Buchdahl} one obtains
$M_S\ge  \left( -9/16 + 81 \arcsin (2\sqrt{2}/3)/(32\sqrt{2})\right)
m_{TB}\approx 1.65 m_{TB}$; that again becomes an equality for
constant density stars.  Constant density stars are the most economic
configurations in the sense that they minimize the ratio $M_S/m_{TB}$
-- and, consequently, the gravitational potential energy $-\tilde U$
-- amongst the subset of static solutions.

\section{Conclusions}

Before pursuing further, one should be aware that the assumption of compactness
of a fluid body made in Sec. 6 imposes  some restrictions on the
polytropic index.   In the case of $\tilde p - \rho $ equations of state the condition
is roughly  $\Gamma >6/5$, but for the $\tilde p - n$  case the situation is  unclear
- the application of existing results concerning the general barotropic models
(\cite{Simon},   \cite{walter}   and \cite{Heinzle})
requires a separate investigation.
 We assume that the polytropic
index is just right to agree with the compactness.

Collecting together the
results of Secs 5 and 6,    we arrive at the
inequality, valid for both $\tilde p - n $ and $\tilde p - \rho $ models,
\begin{eqnarray}
 E_{bind} &\ge  &M_S -{P\over \Gamma -1} -m_{TB}\ge \nonumber \\
&&-{P\over \Gamma -1}+F+{m^2_{TB}\over R}
{2\over \left( 1+\sqrt{1-{2m_{TB}\over R}}\right)^2}  .
\label{6.1}
\end{eqnarray}
A weaker but more elegant than  (\ref{6.1}) is the inequality
\begin{eqnarray}
 E_{bind} &\ge
 -{P\over \Gamma -1}+{3m^2_{TB}\over 5R},
\label{6.2}
\end{eqnarray}
which retains  the same form   in the newtonian limit.
 The sufficient condition that the binding energy be positive is
\begin{eqnarray}
 {P\over \Gamma -1}  < F+
 {m^2_{TB}\over R} {2\over \Bigl( 1+\sqrt{1-{2m_{TB}\over R}}\Bigr)^2}
\label{6.3}
\end{eqnarray}
or, more simply,
$$  {3m^2_{TB}\over 5R} >{P\over \Gamma -1},$$
both  for $\tilde p -  n$
 and $\tilde p -  \rho $ equations of state. An alternative form of a sufficient condition,
$${P\over \Gamma -1}  <
 {M^2_S \over 2L}$$
can be obtained due to another  inequality $M_S-m_{TB}\ge M^2_S/(2L)$ (where $L$
is the geodesic radius  of a compact body), proven in \cite{bmom}.
There exists a simple interpretation of this condition   in terms of the speed of sound.
Namely, define an "average speed of sound $\tilde a$ " by
$P\equiv \tilde a^2 M_S/\Gamma $.
The quantity $\tilde a$   is bounded from below and above by the
minimal $a_{min}$ and maximal $a_{max}$ speeds of sound, respectively,
within the compact body. The formerly
formulated sufficient condition implies now
\begin{equation}
\tilde a^2 <  \Gamma \left( \Gamma -1\right) {3m_{TB}^2 \over 5RM_S}.
\label{6.4}
\end{equation}
  But $M_S\ge m_{TB}$; therefore
\begin{equation}
\tilde a^2< \Gamma \left( \Gamma -1\right) {3m_{TB}  \over 5R }
\label{6.5}
\end{equation}
as the consequence  of    the positivity of the binding energy.
In the newtonian limit $M_S\approx m_{TB}$ and (\ref{6.5}) becomes just another sufficiency
condition for $E_{bind}>0$.
  This    speed of sound  should not be bigger   than (roughly)
the escape velocity (or, strictly saying, the first cosmic velocity $m_{TB}/R$),
if $\Gamma \le 5/3$, and it decreases with the
decrease of $\Gamma $.  And conversely, the inequality $ (\tilde a)^2\ge
\Gamma \left( \Gamma -1\right) {3m_{TB}^2 \over 5RM_S}$ is necessary for
$M_S-{P\over \Gamma -1}-m_{TB}$  being negative,
and thus necessary for the binding energy being negative.

Another form of the necessary condition  can be found for
$\tilde p - \rho $ equations of state. It follows from (\ref{2.4}) that
\begin{equation}
E_{bind}=\int_VdVn-m_{TB} \ge  {M_S \over  (1+{a^2_{max}\over \Gamma })^{1\over \Gamma -1}} -m_{TB}.
\label{6.6}
\end{equation}
The use of the estimate (\ref{5.2a}) of the Theorem allows us to  say that
$E_{bind}$ can be negative only if
\begin{eqnarray}
a^2_{max} &\ge &\Gamma   \left( 1+F+{m_{TB}\over R}{2\over \Bigl( 1+
\sqrt{1-{2m_{TB}\over R}}\Bigr)^2}\right)^{\Gamma -1}
\nonumber \\
&& -\Gamma  ,
\label{6.7}
\end{eqnarray}
or (in a simpler and weaker form)
\begin{eqnarray}
a^2_{max} \ge \Gamma \left[ \left( 1 +{3m_{TB}\over 5R}\right)^{\Gamma  -1}-1\right] .
\label{6.8}
\end{eqnarray}
Since $x\equiv {3m_{TB}\over 5R}\le 4/15$, one has
 $\left( 1+x\right)^\delta \ge
1 +\delta \left( 15\over 19 \right)^{\delta -1}$.
Therefore Eq. (\ref{6.8}) can be
written as
\begin{eqnarray}
a^2_{max} \ge {3\over 5} \Gamma \left( \Gamma -1\right)
\left( {15\over 19}\right)^{2-\Gamma } {m_{TB}\over R}.
\label{6.9}
\end{eqnarray}
This clearly demonstrates, that the negative binding energy requires
the maximal speed of sound to be close to the escape velocity.
For compact objects satisfying the Buchdahl limit we get from (\ref{6.7}),
putting $\Gamma =5/3$, the necessary condition  $a^2_{max}>0.66$.
Let us remark that (\ref{6.8}) and (\ref{6.9}) yield
much worse estimates  $a^2_{max}>0.28$ and $a^2_{max}>0.27$, respectively.
In such a  case  the upper limit for the maximal allowed kinetic energy
 (with $E_{bind}>0$) exceeds  $(\Gamma -1) \times 0.65m_{TB}$ (see the
end of the preceding section);
the kinetic energy can constitute a significant fraction of the mass $m_{TB}$.

By reversing the argument, one obtains from the converse of
(\ref{6.7}) - (\ref{6.9}) a set of  sufficiency conditions for the positivity
of $E_{bind}$. The strongest statement is that if the maximal speed
of sound is smaller than the right hand side of (\ref{6.7}), then
$E_{bind}>0$.

 As is well known,
the negative sign of the binding energy is correlated with the instability
of static fluids \cite{tooper}. Thus  inequalities (\ref{6.4}) and
(\ref{6.7} -  \ref{6.9}) agree well with the
intuitive notion that {\it hot systems} (i. e., with high speed of sound)
are unstable. It is also remarkable that the upper limit increases monotonically
with the increase of $\Gamma $.

We conjecture that inequalities (\ref{6.1} - \ref{6.9})   hold true for systems with
the general barotropic equation $\tilde p = K\rho^{\Gamma (\rho ) }$ (under a suitable definition
of the constants). A plausible form of a related condition in the
case of stationary nonspherical  fluid  bodies
can be  ${P\over \Gamma -1}  <  m^2_{TB}/ \sqrt{S\over \pi }$.

Acknowledgments.    This work has been supported
in part  by the KBN grant 2 PO3B  006 23.

\end{document}